\documentclass[pra,showpacs,twocolumn]{revtex4}
\usepackage{amssymb}
\usepackage{graphicx}

\begin{document}

\title{Three-dimensional transformation for point rotation coordinate
frames. }
\author{Boris V. Gisin}
\affiliation{IPO, Ha-Tannaim St. 9, Tel-Aviv 69209, Israel. E-mail: gisin@eng.tau.ac.il}
\date{\today }

\begin{abstract}
\noindent \noindent We consider the transformation for the point rotation
frames with the angle, spatial coordinate along the axis of rotation and
time as variables. The problem arises when light, propagating through 3-fold
electrooptical crystal, is modulated by the circularly polarized
electromagnetic wave traveling along the optical axis of the crystal. With
help of the transformation we show that such a wave cannot produce an extra
optical frequency shift discussed earlier. In contrast to that the
modulation by the rotating spatially invariable electric field produces the
shift. The formal change to this case may be carried out by reducing the
velocity of the traveling wave to zero. Some properties of the
three-dimensional transformation are discussed.
\end{abstract}

\maketitle


\section{Introduction}

Recently the general linear two-dimensional transformation for point
rotation coordinate frames was considered \cite{jop}. A distinguishing
feature of the frame, in contrast to the Cartesian one, is \ the existence
of the rotation axis at every point. The frame coordinates are an angle and
time; the frequency of rotation is a parameter. The concept of the frame
originates from the optical indicatrix (index ellipsoid) \cite{Nye},\cite%
{Kam}. Rotation of the optical indicatrix arises in three-fold
electrooptical crystals under the action an rotating electric field applied
perpendicular to the optical axis \cite{sm}. Any rotating field also has the
axis of rotation at every point. In this regard the point rotation frame is
more adequate for the description of such fields than the Cartesian frame.

The consideration is motivated by necessity to know the frequency
superposition law. The law is a consequence of the transformation and used
for the description of the single-sideband modulation of light in
electrooptical crystals \cite{sm},\cite{jpc}. The description is based on an
approach connected with the transition to the frame with the resting
indicatrix \cite{sm}. The approach is shortly described below.\ 

Consider a plane circularly polarized light wave propagating through the
crystal with the rotating indicatrix. Transit to a frame with the resting
indicatrix. The frequency of the wave is changed in correspondence with the
law. If the amplitude of the modulating electric field equals the half-wave
value then at the crystal output the circular polarization of the wave\ is
reversed. After the transition to the initial frame the frequency shift is
doubled. If the law is linear one then the shift equals the frequency of the
modulating electric field (for the Pockels crystals). In the opposite case
there must be an extra optical frequency shift. The linear law follows from
the Galilean transformation which is postulated a priori.

In the general transformation the reverse frequency, i.e., the frequency of
the first frame relative to the second one, is a function of the direct
frequency. However both the frequencies are assumed to be symmetric. It
means that the direct frequency is the same function of the reverse
frequency. Using symmetry of the frame coordinates it was shown that three
different types of the transformation are possible. The first type is a
generalization of the Lorentz transformation. The second and third types are
principally different and possess unusual properties, in particular, an
uncertainty of the time determination. In Ref. \cite{jop} an experiment for
testing the second type of transformation was described with the modulation
by the rotating spatially invariable electric field. Such a modulation
corresponds to two-dimensional transformation.

The point rotation frames have no transverse coordinates and, therefore,
they are not compatible with the Cartesian frames. However a coordinate
along the axis of rotation can be used as the spatial coordinate.

In this paper we consider three-dimensional transformation for the point
rotation frames with such a coordinate. For the point rotation frames we
have no a principle, like the relativity principle, for obtaining the exact
form of the transformation. Instead we use some symmetry considerations and
the expansion of the transformation parameters in power series in terms of
the direct and reverse frequency. As it was assumed in \cite{jop} the
characteristic time $\tau $ inherent in the transformation is about "nuclear
time" of the order of $\tau \sim 10^{-23}\sec $. The time is defined as the
size of proton divided by the speed of light. Therefore the normalized
frequency in the optical range is about $10^{-8}\div 10^{-9}$ and we can
restrict the consideration to the first terms of the expansion. Moreover we
assume that parameters connected with the spatial coordinate in the first
approximation correspond values in the Lorentz transformation.

\section{Rotation of the optical indicatrix in 3-fold crystals.}

It is well known that propagation of plane optical waves along the optical
axis of 3-fold crystals is described by the section of the indicatrix
perpendicular to the optical axis \cite{Nye},\cite{Kam} 
\begin{equation}
\frac{x_{1}^{2}+x_{2}^{2}}{n^{2}}%
+2r_{61}x_{1}x_{2}U_{1}+r_{12}x_{1}^{2}U_{2}+r_{22}x_{2}^{2}U_{2}=1,
\label{o3m}
\end{equation}%
where $n$ is the ordinary refractive index, $U_{k}$ is the applied electric
field, $r_{mn}$ are electrooptical coefficients$\,\,r_{22}=-r_{12}=-r_{61}%
\equiv r$ \cite{Nye}.

Using rotation around the optical axis \thinspace\ $x_{1}=x\cos \alpha
-y\sin \alpha ,\,x_{2}=x\sin \alpha +y\cos \alpha $, \thinspace\ we reduce
section (\ref{o3m}) to the principal axes 
\begin{equation}
\frac{x^{2}}{n_{x}^{2}}+\frac{y^{2}}{n_{y}^{2}}=1,\text{ \ }\tan 2\alpha =%
\frac{U_{1}}{U_{2}},  \label{pax}
\end{equation}%
where the refractive indices are given by 
\begin{equation}
\;n_{x,y}\approx n\mp \frac{1}{2}\,n^{3}r\sqrt{U_{1}^{2}+U_{2}^{2}}\,.
\label{eq4}
\end{equation}

We consider the rotation of the optical indicatrix in two important cases.
In the first one the applied electric field rotates around the optical axis
but spatially is invariable: $U_{1}=U\sin \Phi ,$ $U_{2}=U\cos \Phi ,$ where 
$\Phi =\Omega t,$ $\alpha =\Omega t$ $/2,\Omega $ is the frequency of the
electric field. This is achieved by applying a electric field to the system
of electrode pairs on the side faces of crystal. The rotation is provided by
the corresponding phase shift for every pair \cite{jpc},\cite{pat}. The
first case corresponds to infinitely large velocity or infinitely small
length of crystal. Alternatively the time-transit problem must be taken into
account.

In the second case the rotation is created by a powerful traveling
circularly polarized electromagnetic wave with $\Phi =(\Omega t-kz),$ where $%
k$ is the propagation constant, the angle $\alpha =(\Omega t-kz)/2.$ The
optical indicatrix not only rotates but also moves with the velocity $%
u=\Omega /k$. In the second case the phase matching between the modulating
and modulated wave is necessary.

The difference $n_{y}-n_{x}=n^{3}rU$ do not depend on time. If the phase
retardation between two light components along the principal axes of the
indicatrix after passing through the crystal equals $\pi $, i.e., if $%
n^{3}rUd=\lambda /2,$ then the crystal is equivalent to the rotating (or
rotating and moving) half-wave plate. Here $\lambda $ is the light
wavelength, $d$ is the crystal length.

\section{Two-dimensional transformation}

The general linear transformation for the transition from one frame to
another may be written as follows \cite{jop} 
\begin{equation}
\tilde{\varphi}=q(\varphi -\nu t),\;\;\;\tilde{t}=\frac{\tilde{q}q-1}{\tilde{%
q}\tilde{\nu}}\varphi -q\frac{\nu }{\tilde{\nu}}t,  \label{trp}
\end{equation}%
where $\varphi $ is an angle (the angle between the electric vector of a
plane circularly polarized light wave and an axis in the first frame), $t$
is time, $\nu $ is the frequency of second frame relative to first one,
tilde corresponds to the reverse transformation. It is obvious that (\ref%
{trp}) turns out into the reverse transformation if variables with tilde
change to variables without tilde and vice versa.

The functions $\tilde{\nu}(\nu )$ and $q(\nu )\ $remain indeterminate except
the condition at small $\nu $, namely, $\tilde{\nu}\rightarrow -\nu
,\;q\rightarrow 1$ if $\nu \rightarrow 0$. Moreover if $\tilde{\nu}=f(\nu )$
\ then \ $\nu =f(\tilde{\nu}).$ Using new variables $(\tilde{\nu}+\nu )$\
and $\tilde{\nu}\nu ,$ both the equations may be rewritten in a symmetric
form

\begin{equation}
\tilde{\nu}+\nu =F(x),  \label{nugen}
\end{equation}%
where $F$ is a function with the condition $F(0)=0$, $x\equiv \tilde{\nu}\nu 
$ is a parameter. It is assumed that \ $\tilde{\nu}$ and $\nu $ have
opposite signs therefore $x$ is always negative.

In \cite{jop} it was found that, because of symmetry of $\varphi $ and $t$
and the condition (\ref{nugen}), the types of the transformation are defined
by the equation \ 
\begin{equation}
F\left( \frac{\Theta ^{2}}{x}\right) -\frac{\Theta }{x}F(x)=0,  \label{eqnu}
\end{equation}%
where $\Theta =(1-1/q\tilde{q})/\sigma ,$ $\sigma $ is a dimensional
constant.

Three type of solutions of \ Eq. (\ref{eqnu}) exist. First type is exact
solution $\Theta =\tilde{\nu}\nu $. The transformation (\ref{trp}) for the
solution of the first type is a generalization of the Lorentz transformation.

The second type of solutions may be presented as series%
\begin{equation}
\Theta =c\sqrt{-x}+\sum_{n=2}c_{n}(\sqrt{-x})^{n},  \label{qng}
\end{equation}%
where $c=\pm \sqrt{-r},$ $c_{n}$ are constants, $r$ is a real negative root
of the equation $F(r)=0$.\ For multiple roots the first term of expansion (%
\ref{qng}) is proportional to $(-x)^{\frac{1}{2m}}$, where $m$ is
multipleness of the root, with corresponding changes in next terms; in this
paper we consider the most important case with $m=1$. The third type is a
supplement to the second type and corresponds to complex roots.

Examples of solutions of the second and third types are shown in Fig. 1. The
examples represent the function $\Theta (\nu )$ for $F(x)=xP_{n}(x),$ where $%
P_{n}$ is a polynomial of power $n$ with positive coefficients. For such
polynomials solutions of Eq. (\ref{eqnu}) exist only for negative $x$. The
solutions of the second type cross each other at the point $(0,0)$, others
are solutions of the third type. The constant $g=0.01$. Increasing this
constant leads to the increase of asymmetry between the right and left part
of the figure. Obviously the examples are far of exhausting all
possibilities.

\begin{figure}[tbp] 
\vskip+0.1in \centering
\includegraphics[width=1.00\linewidth]{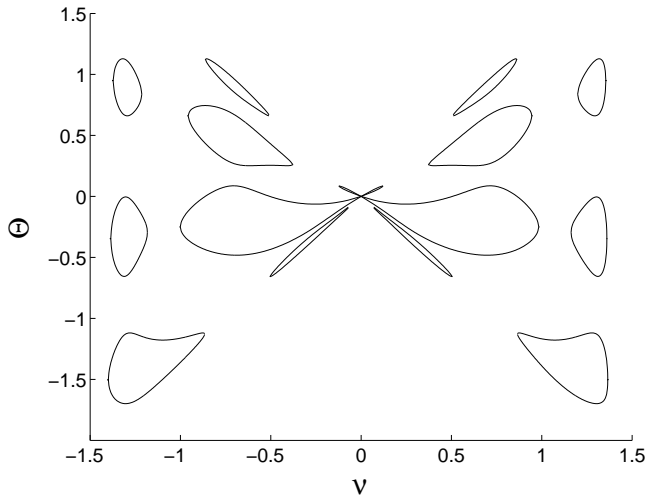} 
\caption{{\small {Solutions of the second and third type for \\
$F=gx[-0.06+(0.5+x)^{2}][0.001+(1.7+x)^{2}]$}}}
\label{f1}
\end{figure}

The expansion of $q(\nu )$ in power series in $\nu $ cannot correspond to
the solution (\ref{qng}). Such a correspondence is possible if the expansion
has the form 
\begin{equation}
q=\sum_{n=0}(a_{n}+\varepsilon a_{n}^{\prime })\nu ^{n},  \label{qs}
\end{equation}%
where $\varepsilon \equiv \pm \sqrt{-x}/\nu ,$ $\tilde{\varepsilon}=\pm 
\sqrt{-x}/\tilde{\nu},$ for $\nu \rightarrow 0$ $\varepsilon \rightarrow \pm
1,\tilde{\varepsilon}\rightarrow -\varepsilon .$ In this case we do not miss
the second and third types of solutions. Unfortunately in Ref. \cite{jop}
the value $+\varepsilon $ was used for $\tilde{\varepsilon}.$ Therefore the
mistaken conclusion was made about asymmetry of the extra optical frequency
shift by reversing of the electric field rotation. However the schematic of
the experiment proposed in the paper \cite{jop} allows to measure as
asymmetric as symmetric shifts.

The transformation \cite{sm} for the solution of the second type possesses
an unusual property. For $\nu \rightarrow 0$ one follows from Eq. (\ref{qng}%
) \ $(1-1/q\tilde{q})/\tilde{\nu}\rightarrow \varepsilon \tau $. In this
limit Eq. (\ref{trp}) is%
\begin{equation}
\tilde{\varphi}=\varphi ,\text{ \ \ }\tilde{t}=\tau \varphi +t.  \label{lt}
\end{equation}%
The limit corresponds to the transformation "into itself". This result is
interpreted as uncertainty of the time determination \cite{jop}. Indeed,
different ways are possible to come in the point (0,0) along one closed
curve in Fig. 1.

Normalizing $\varphi \rightarrow \varphi \sqrt{-\tilde{\nu}/\nu },$ \ \ \ $%
q\rightarrow -q\tilde{\nu}/\nu ,$\ \ \ we arrive to the case $\tilde{\nu}%
=-\nu ,$ $|\varepsilon |=1$. It is known that the Galilean as well as the
Lorentz transformation is invariant under the sign change of both the
velocity and time. Here we extend this principle to the transformation (\ref%
{trp}) and assume that the normalized transformation (\ref{trp}) is
invariant under the change $\nu \rightarrow -\nu ,$ $t\rightarrow -t,$ $%
\tilde{t}\rightarrow -\tilde{t}$. It means that $\tilde{q}=q$ or in the
expansion (\ref{qs}) coefficients $a_{n}$ and $a_{n}^{\prime }$ with odd and
even $n$ respectively are equal zero.

In \cite{jop} it was shown that the extra frequency shift at the crystal
output may be measured only for the second type of solutions. In the first
approximation the shift doesn't depend on $\nu $, i.e., such a shift may be
produced by the usual half-wave plate. It, however, doesn't mean that the
shift could be doubled by passing the second modulator or the half-wave
plate. In this case the polarization of the wave would turn back to its
initial state and the extra shift would disappear.

\section{Three-dimensional transformation}

The transformation in the 3-dimensional case may be written in the following
general form%
\begin{eqnarray}
\tilde{\varphi} &=&q_{1}[\varphi +p_{1}(z-ut)-\nu t],\text{ \ \ \ \ } 
\nonumber \\
\tilde{z} &=&q_{2}[p_{2}(\varphi -\nu t)+z-ut],\text{ \ \ \ \ \ }
\label{eq1} \\
\tilde{t} &=&q_{31}\varphi +q_{32}z+q_{33}t,\text{ \ \ \ \ \ \ }  \nonumber
\end{eqnarray}%
where $\varphi ,\nu $ is defined as in Section II, z is the coordinate along
the axis of rotation, $u$ is the velocity of the second frame (the moving
and rotating optical indicatrix) relative to the first one, all parameters
are functions of $\nu ,u.$ For definiteness we assume that the right-hand
polarized light, propagating in the positive direction of the z-axis, has
positive $u,$ $\nu .$ The terms $p_{1}(z-ut)$ and $p_{2}(\varphi -\nu t)$
describe the velocity and frequency dismatch between the modulating and
modulated wave. Note that using the transformation (\ref{eq1}) in another
form, with the frequency $\nu \prime =\nu +p_{1}u$ and the velocity $u\prime
=u+p_{2}\nu ,$ leads to erroneous results.

Using Eq. (\ref{eq1}) we may find the output frequency in the experiment of
Ref. \cite{jop}. Let \ $\omega =\varphi /t$ and $V=z/t$\ to be the frequency
and velocity of the light wave in the first frame. If the optical frequency
in the second frame $\tilde{\omega}=\tilde{\varphi}/\tilde{t}$\ changes the
sign $\tilde{\omega}\rightarrow -\tilde{\omega}$ then the output frequency
in the initial frame is 
\begin{equation}
\omega _{out}=\frac{\omega -2\tilde{q}_{1}q_{1}\Delta }{1-2\tilde{q}%
_{31}q_{1}\Delta },  \label{vf1}
\end{equation}%
where $\Delta =\omega +p_{1}(V-u)-\nu $.

Now we assume that $\tilde{u}=-u,$ since $u$ is the velocity of rectilinear
move, and normalize the transformation so that $\tilde{\nu}=-\nu $: 
\[
\text{\ }\nu \rightarrow \sqrt{-\tilde{\nu}\nu },\text{ \ \ }(\varphi ,\text{
}p_{1})\rightarrow -\frac{\tilde{\nu}}{\nu }(\varphi ,p_{1}), 
\]%
\[
\text{\ }(q_{1},p_{2},q_{31})\rightarrow (q_{1},p_{2},q_{31})\sqrt{-\frac{%
\nu }{\tilde{\nu}}}. 
\]%
We assume that Eq. (\ref{eq1}) is invariant under the change $\nu =-\nu ,$ $%
u\rightarrow -u,$ $t=-t,$ $\tilde{t}=-\tilde{t}.$ In this case $\tilde{q}%
_{31}=-q_{31},$ $\tilde{q}_{32}=-q_{32},$ whereas other parameters with the
tilde equals to that without the tilde.

Using the reverse transformation, we obtain a system of five independent
equations for seven parameters of Eq. (\ref{eq1}). Expressing the parameters
in the terms of $q_{1},q_{2}$ we find 
\begin{eqnarray}
p_{1} &=&\frac{\nu }{u}\frac{(q_{2}-1)}{q_{1}},\text{ \ }p_{2}=\frac{u}{\nu }%
\frac{(q_{1}-1)}{q_{2}},  \label{pfq} \\
q_{31} &=&-\frac{(q_{1}-1)(q_{1}+q_{2})}{(q_{1}+q_{2}-1)\nu },  \label{q31}
\\
q_{32} &=&-\frac{(q_{2}-1)(q_{1}+q_{2})}{(q_{1}+q_{2}-1)u},  \label{q32} \\
q_{33} &=&(q_{1}+q_{2}-1).  \label{d33}
\end{eqnarray}

Expand $\ q_{1},$ $q_{2}$ in power series in $\nu $ . It is reasonable to
assume that coefficients of the expansion are even functions of $u$ 
\begin{eqnarray}
q_{1} &=&a+\varepsilon a_{1}\nu +a_{2}\nu ^{2}+\varepsilon a_{3}\nu
^{3}\ldots ,\text{ \ }  \label{q1e} \\
q_{2} &=&b+\varepsilon b_{1}\nu +b_{2}\nu ^{2}+\varepsilon b_{3}\nu
^{3}\ldots  \label{p2e}
\end{eqnarray}%
In the limiting case at $\nu \rightarrow 0$ the parameter $q_{31}$ must be
bounded therefore $a=1.$ Moreover $b$ must be equal the corresponding
Lorentz value:\ $b=1/\sqrt{1-\gamma ^{2}},$ where $\gamma =u/c$, where $c$
is the speed of light in crystal. Using the values we obtain in the first
approximation 
\begin{eqnarray}
p_{1} &=&\frac{\nu \gamma }{c\sqrt{1-\gamma ^{2}}(1+\sqrt{1-\gamma ^{2}})},
\\
p_{2}=u &&\varepsilon a_{1}\sqrt{1-\gamma ^{2}},  \label{pb1} \\
q_{31} &=&-\varepsilon a_{1}(\sqrt{1-\gamma ^{2}}+1), \\
q_{32} &=&-\frac{\gamma }{c\sqrt{1-\gamma ^{2}}},\text{ \ \ \ }q_{33}=\frac{1%
}{\sqrt{1-\gamma ^{2}}}.  \label{abc1}
\end{eqnarray}%
With these definitions the transformation Eq. (\ref{eq1}) is%
\begin{eqnarray}
\tilde{\varphi} &=&\varphi +\frac{\nu \gamma }{c\sqrt{1-\gamma ^{2}}(1+\sqrt{%
1-\gamma ^{2}})}(z-ut)-\nu t,  \label{phi} \\
\tilde{z} &=&u\varepsilon a_{1}(\varphi -\nu t)+\frac{z-ut}{\sqrt{1-\gamma
^{2}}},  \label{zp} \\
\tilde{t} &=&-\varepsilon a_{1}(\sqrt{1-\gamma ^{2}}+1)\varphi +\frac{%
-\gamma z/c+t}{\sqrt{1-\gamma ^{2}}}.\text{ \ \ \ }  \label{tp}
\end{eqnarray}%
In the condition of the phase matching $z=ct,$ $u\rightarrow c$ the reverse
velocity $\tilde{z}/\tilde{t}$ must have physically acceptable values. From
this requirement one follows that $a_{1}$ may be presented in the form $%
a_{1}=-\frac{1}{2}(\sqrt{1-\gamma ^{2}})^{\xi }\tau (\gamma ),$ where $\xi
\geq 1,$ $\tau (\gamma )$ is a function, $\tau (0)$ equals the introduced
above characteristic time $\tau $.

Using Eq. (\ref{vf1}) we obtain for two cases $u=0$ and $u=c$ respectively 
\begin{equation}
\omega _{out}=\frac{-\omega +2\nu }{1-2\tau \omega },\text{ \ }\omega
_{out}=-\omega +2\nu \text{\ },  \label{omf}
\end{equation}%
where we neglected small terms with $\nu $ in the first expression. The
output frequency for $u=0$ coincides with the value calculated for the
two-dimensional case. For $u=c$ the extra shift vanishes. Note that this
result coincides for both the normalized and non-normalized transformation.

\section{Conclusion}

We have considered the three-dimensional transformation for the point
rotation frames in two important cases. In spite the fact that there is no a
general physical principle, the transformation may be defined with help of
some symmetry considerations and the expansion of parameters in terms of the
frequency. A surprising result is the absence of the extra shift in the case
of the modulation by traveling electromagnetic wave in condition of the
phase matching.

However the main question about the existence of the extra shift in reality
remains to be open. Some arguments in favour of the positive answer on this
question are below.

The optical indicatrix as a coordinate frame have been used for more than
hundred years. The Galilean transformation for such a rotating frames seems
unacceptable since fields with infinitely large frequencies are not known in
physics. The Lorentz transformation seems too simple because it have only
one limiting frequency. In contrast to that the transformation corresponding
to the solutions of the second and third type possesses a variety
possibilities. The arguments in some extend confirm the necessity to perform
measurements of the proposed extra shift and the characteristic time.

\end{document}